\documentclass[prl,twocolumn,superscriptaddress]{revtex4}
\usepackage{amssymb,amsmath,amsthm,graphicx}
\usepackage{latexsym}
\usepackage{bm}
\usepackage[]{hyperref}
\usepackage{cleveref}
\usepackage{tabularx}

\pdfoutput=1

\begin{document}

\title{Multi-oscillating Boson Stars}

\author{Matthew Choptuik}
\email{choptuik@physics.ubc.ca}
\affiliation{Department of Physics and Astronomy, University of British Columbia, 6224 Agricultural Road, Vancouver, B.C., V6T 1Z1, Canada}
\author{Ramon~Masachs}
\email{rmg1e15@soton.ac.uk}
\affiliation{STAG research centre and Mathematical Sciences, University of Southampton, Highfield Southampton SO17 1BJ, UK}
\author{Benson~Way}
\email{benson@phas.ubc.ca}
\affiliation{Department of Physics and Astronomy, University of British Columbia, 6224 Agricultural Road, Vancouver, B.C., V6T 1Z1, Canada}

\begin{abstract}
We propose that stable boson stars generically fall within an infinite-parameter family of solutions that oscillate on any number of non-commensurate frequencies. We numerically construct two-frequency solutions and explore their parameter space. These solutions merge with the standard boson star family in the limit where the non-dominating frequencies are turned off.  We find that for a fixed energy, these two-frequency solutions can differ considerably in size from standard boson stars.
\end{abstract}

\maketitle

{\bf~Introduction --} Boson stars are compact self-gravitating objects made of a massive complex scalar field. They were first introduced by Kaub and also Ruffini and Bonazzola in the late 1960's \cite{Kaup:1968zz,Ruffini:1969qy}. (See \cite{Liebling:2012fv} for a review.) Besides their intrinsic interest as self-gravitating solitonic solutions in general relativity, they are used in models of gravitational collapse, dark matter, and as gravitationally compact objects. Part of their utility stems from the fact that complex scalar fields do not suffer from issues like shocks and discontinuities that affect fluid dynamics. Thus, relative to neutron stars, for example, boson stars are easier to treat computationally, but still remain useful probes of strong field gravity.

The recent success of the gravitational wave observatories LIGO and VIRGO \cite{Abbott:2016blz,LIGOScientific:2018mvr} has led to a growing interest in the merger dynamics of exotic compact objects like boson stars.  It is important, therefore, to identify the generic configuration of these compact objects that forms from some dynamical process.  In contrast to black holes, there are no uniqueness theorems for boson stars.  Indeed, numerical simulations that do not form black holes or disperse to infinity appear to either leave a stable boson star or approach some oscillating solution \cite{Balakrishna:1997ej,Hawley:2000dt,Hawley:2002zn,Lai:2007tj}. Are these oscillating solutions transient or do they remain indefinitely?  If they are long-lasting, how large is the space of such oscillating solutions?

We propose that such oscillating solutions constitute an infinite dimensional family.  This family includes the usual boson star family, as well as an infinite-parameter space of configurations that oscillate indefinitely on any number of frequencies.  By analogy with similar solutions found in anti-deSitter space~\cite{Basu:2010uz,Dias:2011ss,Maliborski:2013jca,Horowitz:2014hja,Choptuik:2017cyd,Choptuik:2018ptp}, we call these `multi-oscillators'. We will explicitly construct such configurations with two oscillations (double-oscillators), but our methods can in principle be used for including more oscillations.

We mention that in \cite{Hawley:2002zn} strong numerical evidence was presented for the existence of solutions to the Einstein-Klein Gordon system with more than one scalar field. In that case a time evolution with initial data composed of a complex scalar field with the imaginary part phase-shifted was performed and it was found that the system approached a solution that was called a {\emph{phase-shifted boson star}}. Such solutions should lie within the family of multi-oscillators.

Let us now review the phase space and stability of boson stars\cite{Lee:1986ts,Friedberg:1986tp,Lee:1988av,Gleiser:1988rq,Gleiser:1988ih,Ferrell:1989kz,Jetzer:1991jr,Lee:1991ax,Seidel:1990jh,Balakrishna:1997ej,Hawley:2000dt,Hawley:2002zn,Lai:2007tj}.  For concreteness, we consider the theory of a complex scalar $\varphi$ with mass $\mu$, minimally coupled to gravity. Boson stars are derived from an ansatz for the scalar field of the form $\varphi = e^{i\omega_1 t}\psi(r)$ for some real function $\psi$ with time coordinate $t$, and radial coordinate $r$. That is, the complex scalar is spherically symmetric and has periodic time dependence with frequency $\omega_1$. This time dependence only appears as an overall phase, so the equations of motion are independent of $t$ and reduce to a set of ordinary differential equations (ODEs) in $r$.  Solutions to these equations can be parametrised by the frequency $\omega_1$.  Since there is only one periodic oscillation, boson stars are single-oscillators in the multi-oscillator family.

\begin{figure}[h]
\includegraphics[width=.4\textwidth]{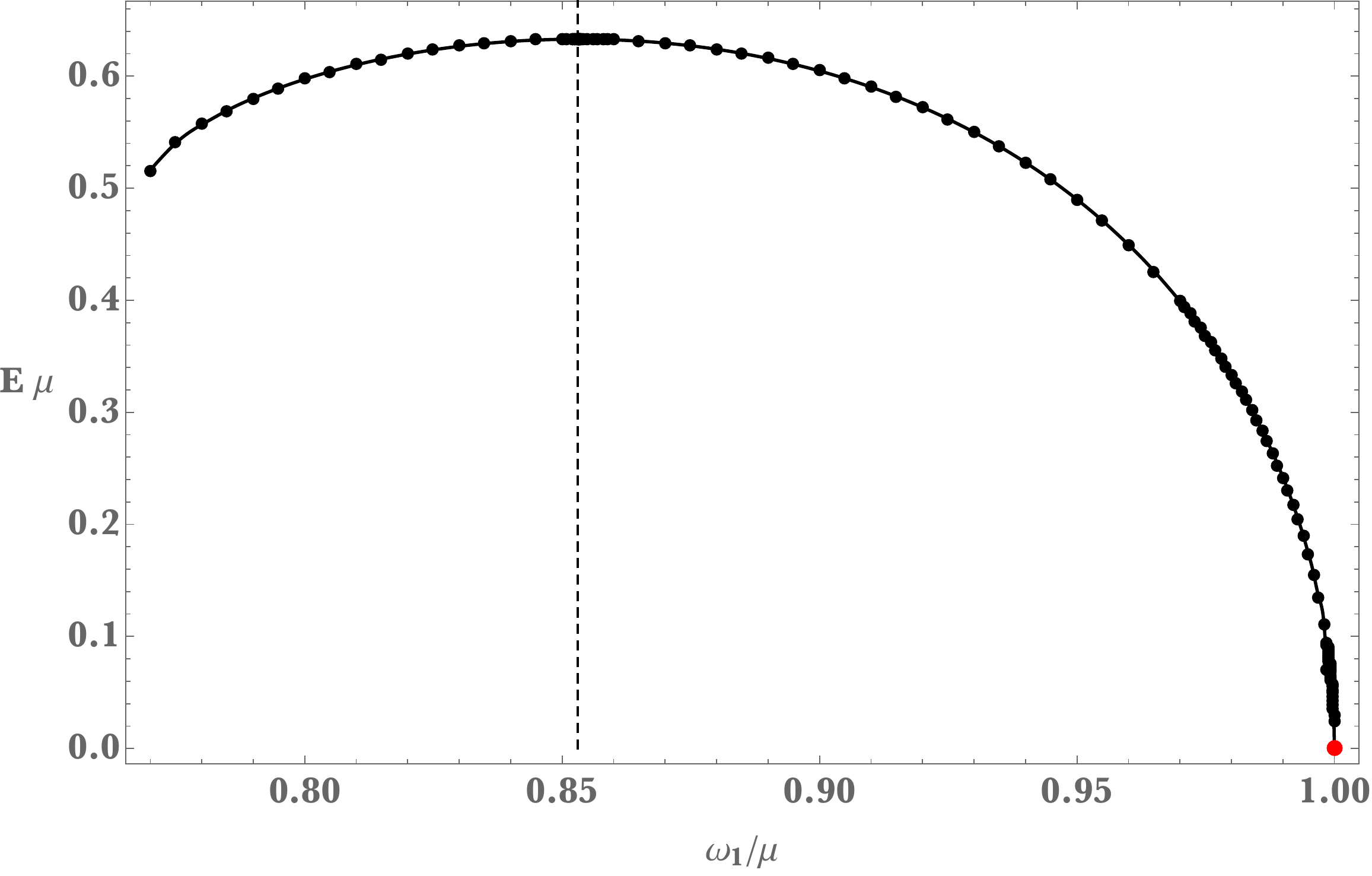}
\caption{Energy of boson stars versus their frequency. The red dot with zero energy marks Minkowski space. The vertical dashed line separates the solutions into the stable branch (right) and the unstable (left).}\label{fig:ebosonstar}
\includegraphics[width=.4\textwidth]{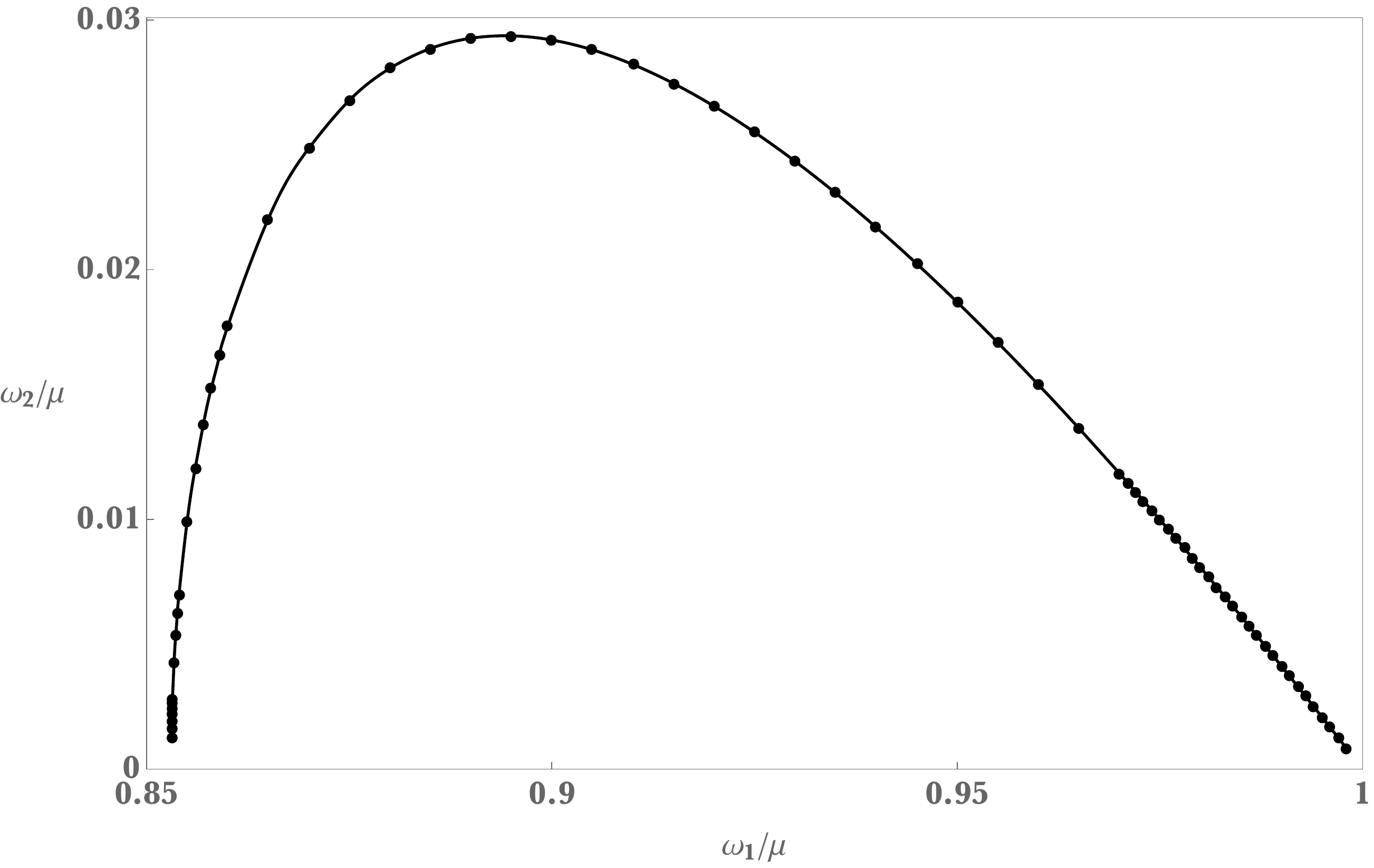}
\caption{Perturbative frequencies of the stable boson star branch. The zero-frequency at $\omega_1/\mu\approx 0.853$ agrees with the maximum energy in Fig.~\ref{fig:ebosonstar}.}\label{fig:perturbations}
\end{figure}

Fig.~\ref{fig:ebosonstar} shows the energy of boson stars versus the frequency $\omega_1$.  We see that there is a maximum energy that divides the solution into two branches. Solutions on the right branch (from arbitrarily small energies up to the maximum energy) are stable, while solutions on the left branch are unstable.

The spectrum of linear perturbations of solutions on the stable branch consists of an infinite number of normal modes.  The lowest normal mode frequency, which we call $\omega_2$, is shown as a function of $\omega_1$ in Fig.~\ref{fig:perturbations}.  The point where $\omega_2$ vanishes (with $\omega_1\sim 0.85$) coincides with the boson star with maximum energy. That is, at this point, $\omega_2$ corresponds to the zero mode that marks the onset of instability of boson stars.

We will demonstrate that a linear normal mode perturbation of a boson star can in fact be extended to a new nonlinear solution. Nonlinear corrections will modify the perturbative frequency $\omega_2$, so this new solution can be parametrised by $\omega_1$ and $\omega_2$. This solution oscillates on both of those frequencies: a double-oscillator.  But the same process can be applied using higher modes than $\omega_2$, as well as any combination of those modes.  This generates a family of solutions that oscillate on any number of frequencies $\omega_i$: a multi-oscillator.

\textbf{Numerical Construction --} We now present the details of our numerical construction of double-oscillators. Our ansatz is
\begin{subequations}\label{ansatz}
\begin{align}
\mathrm ds^2&=-\delta \, f {\mathrm d t^2}+\frac{\mathrm d r^2}{f}+r^2\mathrm d \Omega_2,\label{metricansatz}\\
\varphi&=e^{i \omega_1{t}} \left(\varphi_r+i \varphi_i\right).\label{scalarfield}
\end{align}
\end{subequations}
We use the coordinate $\rho$ given by $r=r_s\rho  \sqrt{2-\rho ^2}/(1-\rho ^2)$ so that the domain $\rho\in(0,1)$ is compact, with the origin at $\rho=0$ and asymptotic infinity at $\rho\to1$ . Here, $r_s$ is a scaling parameter that can be chosen freely without affecting the physical solution. We also treat the mass of the scalar $\mu$ as an overall scale, so it will not appear in our equations. (This is equivalent to setting $\mu=1$). For convenience, take
\begin{align}
\begin{split}
\delta&=1-\left(1-\rho ^2\right)f_1,\\
f&=1-\rho ^2 \left(2-\rho ^2\right) \left(1-\rho ^2\right)f_2,\\
\varphi_r&=\left(1-\rho ^2\right)^2f_3,\\
\varphi_i&=\left(1-\rho ^2\right)^2f_4,
\end{split}
\end{align}
and take $f_i,\ i=1,2,3,4$ to be real functions of $t$ and $\rho$. The equations of motion contain two first-order spatial constraints for $f_1$ and $f_2$, and the two second-order Klein-Gordon equations for the scalar fields $f_3$ and $f_4$. There is also a single first-order temporal constraint equation that we do not solve directly, but monitor as a measure of numerical accuracy.  Note that since $\omega_1$ only appears as an overall phase, the time dependence of the equations of motion lies only in the functions $f_i$ and the derivative $\partial_t$.

The required boundary conditions are regularity at the origin $\rho=0$ ($r=0$) and asymptotic flatness at infinity $\rho\to1$ ($r\to\infty$).  One can also supply initial data for $f_3$ and $f_4$ and their time derivatives at a fixed time, say $t=0$ and solve this system as an initial value problem. For our purposes, however, we wish to find solutions with some specified quasiperiodic behaviour in time. That is, we seek solutions that oscillate with some superposition of periods. We will therefore require additional boundary conditions on $t$.

A quasiperiodic function $f$ on $k$ frequencies has a spectral expansion
\begin{equation}
    f(t,\rho) = \sum_{n_1,\ldots,n_k} A_{n_1,\ldots,n_k}(\rho) e^{in_1\omega_1 t+\ldots+in_k\omega_k t}\;,
\end{equation}
which contains the same spectral information as
\begin{equation}
    f(t_1,\ldots,t_k,\rho) = \sum_{n_1,\ldots,n_k} A_{n_1,\ldots,n_k}(\rho) e^{in_1\omega_1 t_1+\ldots+in_k\omega_k t_k}\;.
\end{equation}
Note that $\partial_t$ and $\partial_{t_1}+\ldots+\partial_{t_k}$ are equivalent.  However, the functions themselves are not the same under $t\to t_1+\ldots +t_k$, so we require that the equations of motion are independent of $t$, except for appearances of the derivative $\partial_t$ and the function $f$.

Multi-oscillators can therefore be found by setting $\partial_t\to \partial_{t_1}+\ldots+\partial_{t_k}$ in the equations of motion, promoting the functions $f_i(t,\rho)\to f_i(t_1,\ldots,t_k,\rho)$, and demanding each coordinate $t_i$ to be periodic with frequency $\omega_i$.  In general, the result of this process defines a boundary value problem on $k+1$ coordinates which can be solved numerically.

We have somewhat simplified this process in our ansatz by placing one of the frequencies $\omega_1$ into an overall phase, removing the time dependence in $t_1$.  Single-oscillators (boson stars) can therefore be found setting $f_4=0$ and removing any time dependence in the functions, leading to a set of ODEs in $\rho$. Double oscillators can be found by allowing the $f_i$ to be independent of $t_1$ and periodic on the coordinate $t_2$ with frequency $\omega_2$, resulting in a partial differential equation (PDE) in $t_2$ and $\rho$.

Without loss of generality, we can choose the time Fourier series of $f_1$, $f_2$, and $f_3$ to be cosine series, and $f_4$ to be a sine series.
\begin{align}
    f_i(t_2,\rho)&=\sum_k\hat f_{i}^{(k)}(\rho)\cos(k\omega_2 t_2)\qquad i\in\{1,2,3,\}\;,\nonumber\\
    f_4(t_2,\rho)&=\sum_k\hat f_{4}^{(k)}(\rho)\sin(k\omega_2 t_2)
\end{align}
We introduce the Fourier coefficients
\begin{align}
\epsilon_1=\hat f_3^{(0)}(0),\qquad \epsilon_2=\hat f_4^{(1)}(0),
\end{align}
which essentially measure the amplitudes of the corresponding fields $f_3$ and $f_4$ at the origin. We find $\epsilon_2$ to be more convenient than $\omega_2$ as a parameter since at $\epsilon_2=0$ one recovers the boson star.  We can therefore treat $\epsilon_2$ as a measure of our deformation from the boson star solution.  Thus we use the parameters $\{\omega_1,\epsilon_2\}$ to move numerically in phase space.  We take $r_s \mu=5/\sqrt\epsilon_1$ as a convenient choice of scaling parameter $r_s$.

The main features we extract from our numerical calculation are the frequencies $\omega_1$ and $\omega_2$, as well as the energy $E$ and the mass aspect function $M$:
\begin{align}
\frac{E}{\mu}=\frac{r_s}{2}\bar{f_2}(1),\quad \frac{M(\rho)}{\mu}=\frac{r_s\rho^3\left(2-\rho^2\right)^{3/2}}{2}\bar{f_2}(\rho).
\end{align}
Notice that the mass aspect function tends to the energy as $\rho\to 1$. Additionally, since the solutions are periodic in time, we compute these quantities by taking the average over a period in $t_2$ (which we express using a bar, as in $\bar{f_2}$) at a fixed radius $\rho$.  Though we've taken $E$ to be an average over a period of $t_2$, energy conservation actually guarantees that ${f_2}(t_2, 1)$ is constant. We can therefore use the standard deviation of ${f_2}(t_2, 1)$ as a check on numerics.

We solve the double-oscillator equations numerically using Fourier spectral methods in $t_2$, and fourth order finite differences in $\rho$.  We use a Newton-Raphson method with the boson star solutions as initial estimates.  For our numerical algorithm, we used 31 gridpoints in the time direction and 71 in the radial direction. For data shown here, the temporal constraint and the standard deviation of ${f_2}(t_2, 1)$ over a period are smaller than $10^{-6}$.

\textbf{Results --}
As we have mentioned earlier, solutions with small $\epsilon_2$ are well-approximated by linear perturbation theory about boson stars, where $\omega_2$ is the perturbative normal mode frequency, which have already been shown in Fig.~\ref{fig:perturbations}.

\begin{figure}
\centering
\includegraphics[width=.4\textwidth]{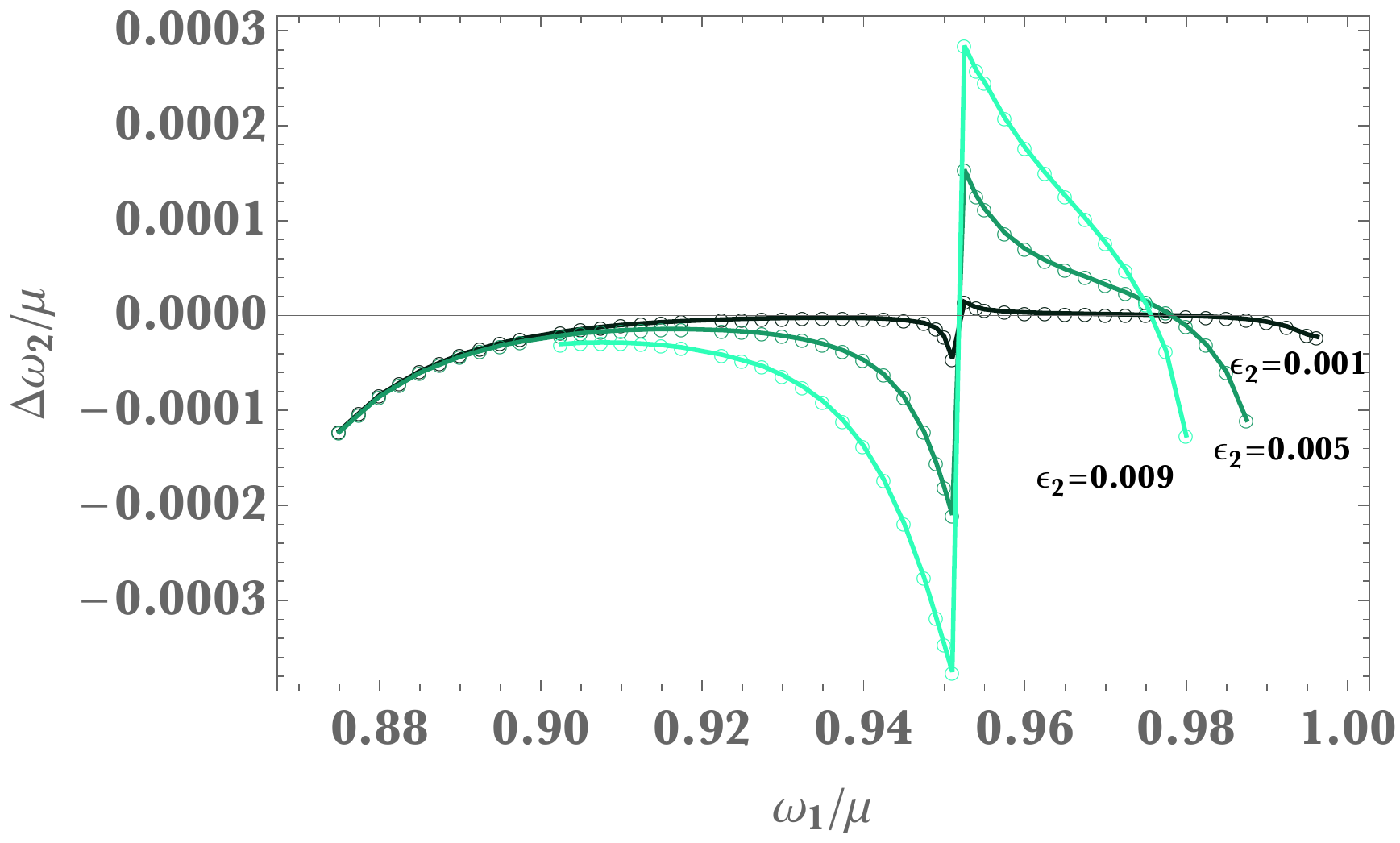}
\caption{Change in frequency $\omega_2$ with respect to the normal mode of the boson star as a function of $\omega_1$.}\label{fig:smallepsfreqdiag}
\end{figure}

In Fig.~\ref{fig:smallepsfreqdiag}, we show how $\omega_2$ changes from the perturbative boson star value ($\Delta \omega_2=\omega_2-\omega^{(\rm BS)}_2$) as $\epsilon_2$ is increased. Depending on $\omega_1$, $\omega_2$ may either increase or decrease from its perturbative value. 
Around $\omega_1\sim0.952$ there appears to be a divergence in this figure, which arises as a result of a degeneracy of normal modes.

We mention that a similar divergence has been observed in toroidal perturbations of black branes \cite{Dias:2017coo}. In the black brane, this divergence is a consequence of a particular alignment of perturbations.  The usual perturbations are ill-defined at this divergence, and are replaced by a special and distinct set of perturbations with different symmetry properties.  In the present double-oscillator case, there may likewise be a special double-oscillator generated by a distinct set of perturbations, but such a solution would not be generic.

Despite this feature of a divergence, we conclude from this figure that the secondary frequency $\omega_2$ does not typically differ too far from the perturbative value.  This could be anticipated from the fact that corrections to the frequency $\omega_2$ occur at higher orders in perturbation theory.

We now compare various quantities between boson stars and multi-oscillators at fixed energy. We normalize the energy with respect to the maximum energy of the boson star shown in Fig.~\ref{fig:ebosonstar} using $E/E^{(\rm BS)}_{\rm max}$. We note that all of the solutions we have found satisfy $E/E^{(\rm BS)}_{\rm max}\leq1$. It is conceivable that some multi-oscillators may have higher energy than the maximum boson star energy, but these would most likely exist close to the critical frequency $\omega_1=0.853$, where finding such solutions is numerically challenging.

\begin{figure}
\centering
\includegraphics[width=.4\textwidth]{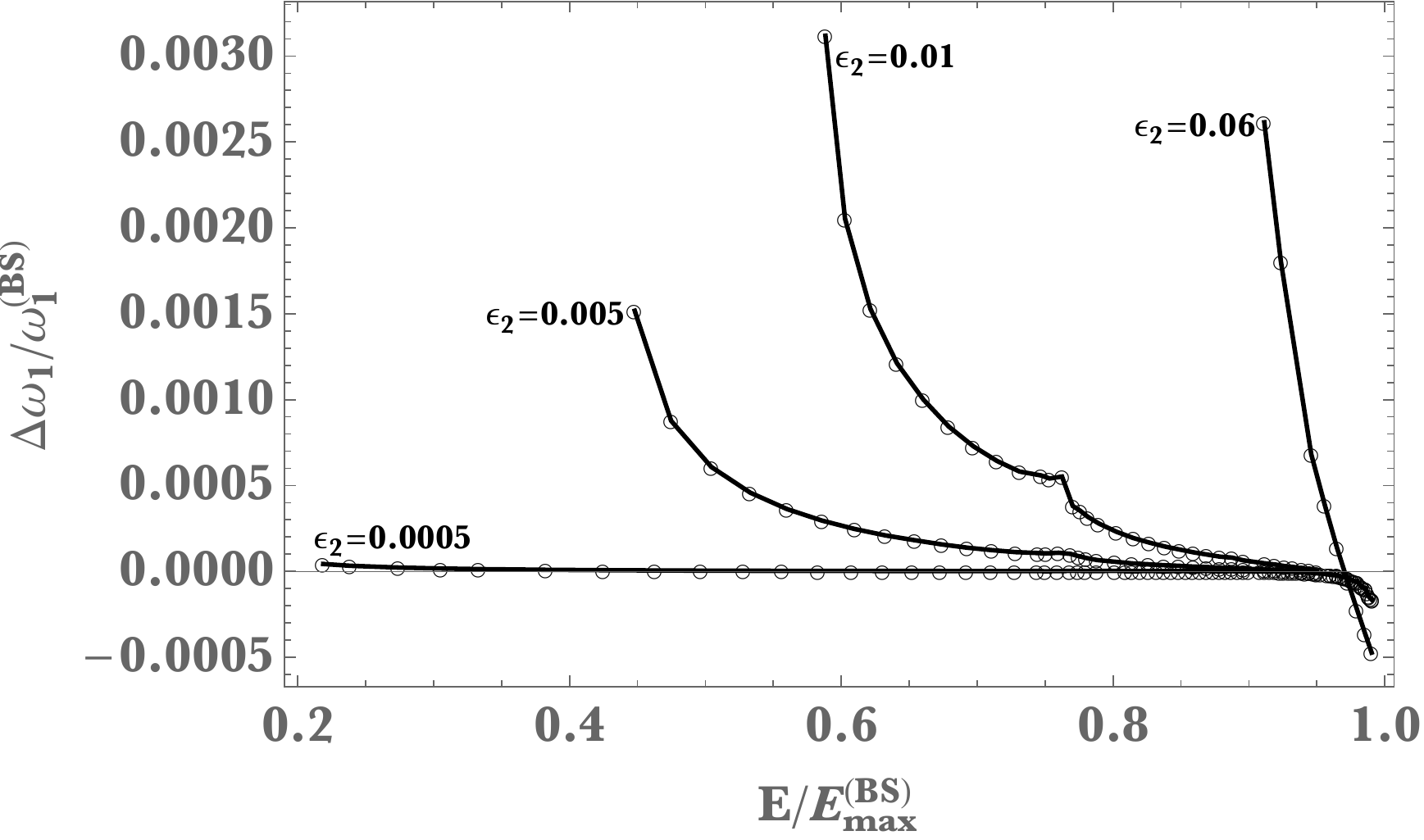}
\includegraphics[width=.4\textwidth]{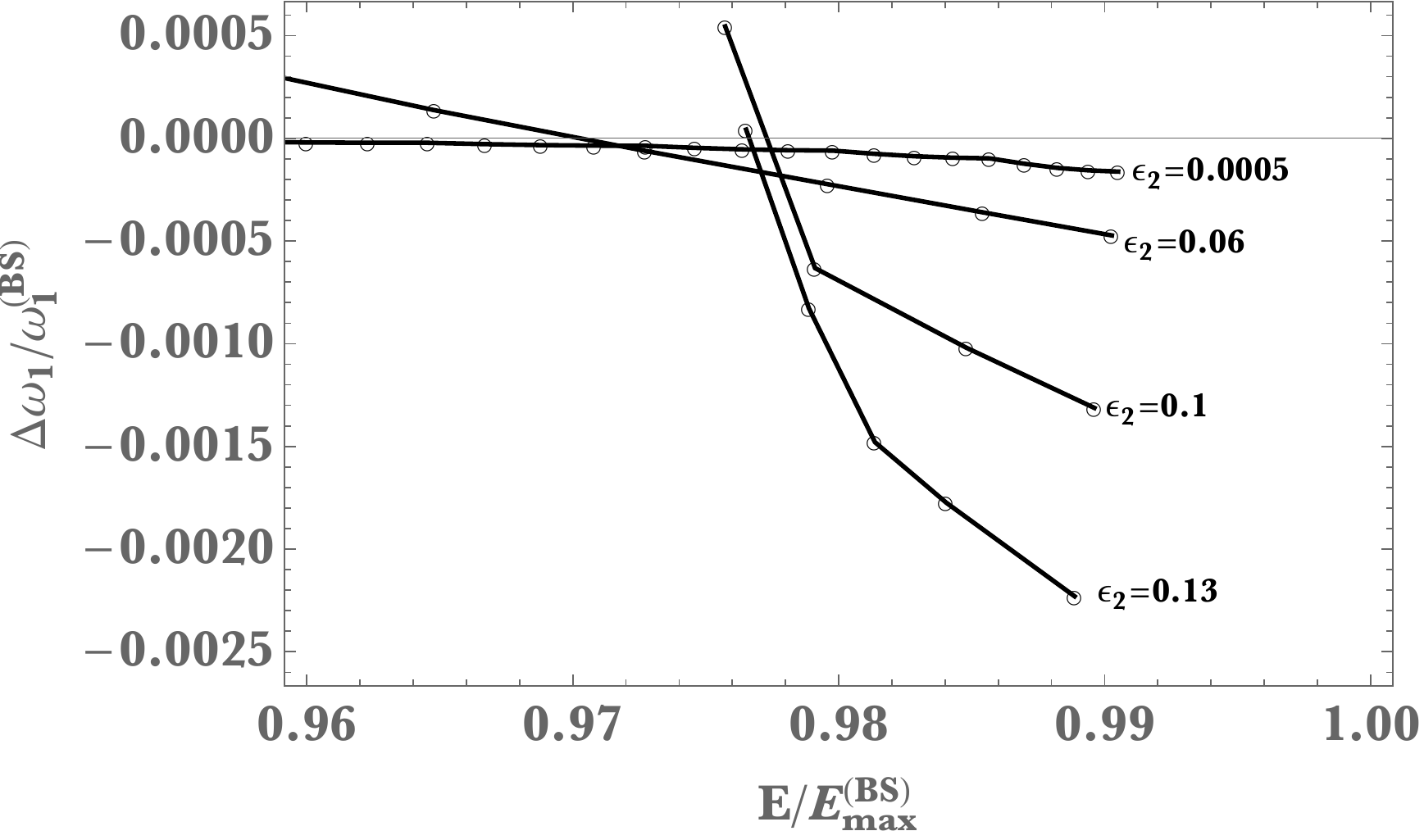}
\caption{Difference in principal frequency $\Delta\omega_1\equiv\omega_1-\omega^{(\rm BS)}_1$ as a function of the energy. {\emph{Top:}} larger range of energies, in particular the line with $\epsilon_2=0.0005$ goes from frequencies $\omega_1=0.997$ to $\omega_1=0.875$. {\emph{Bottom:}} Zoom into the region of larger energies where we have obtained solutions for larger values of $\epsilon_2$. }\label{fig:o1E}
\end{figure}

In figure \ref{fig:o1E} we plot the relative difference between the multi-oscillator primary frequency and the frequency of the boson star $\Delta\omega_1/\omega^{(\rm BS)}_1\equiv\omega_1/\omega^{(\rm BS)}_1-1$ as a function of the energy.  As was seen for $\omega_2$, we find that $\omega_1$ can either decrease or increase from the boson star solution with the same energy, but the difference tends to remain small.

\begin{figure}
\centering
\includegraphics[width=.4\textwidth]{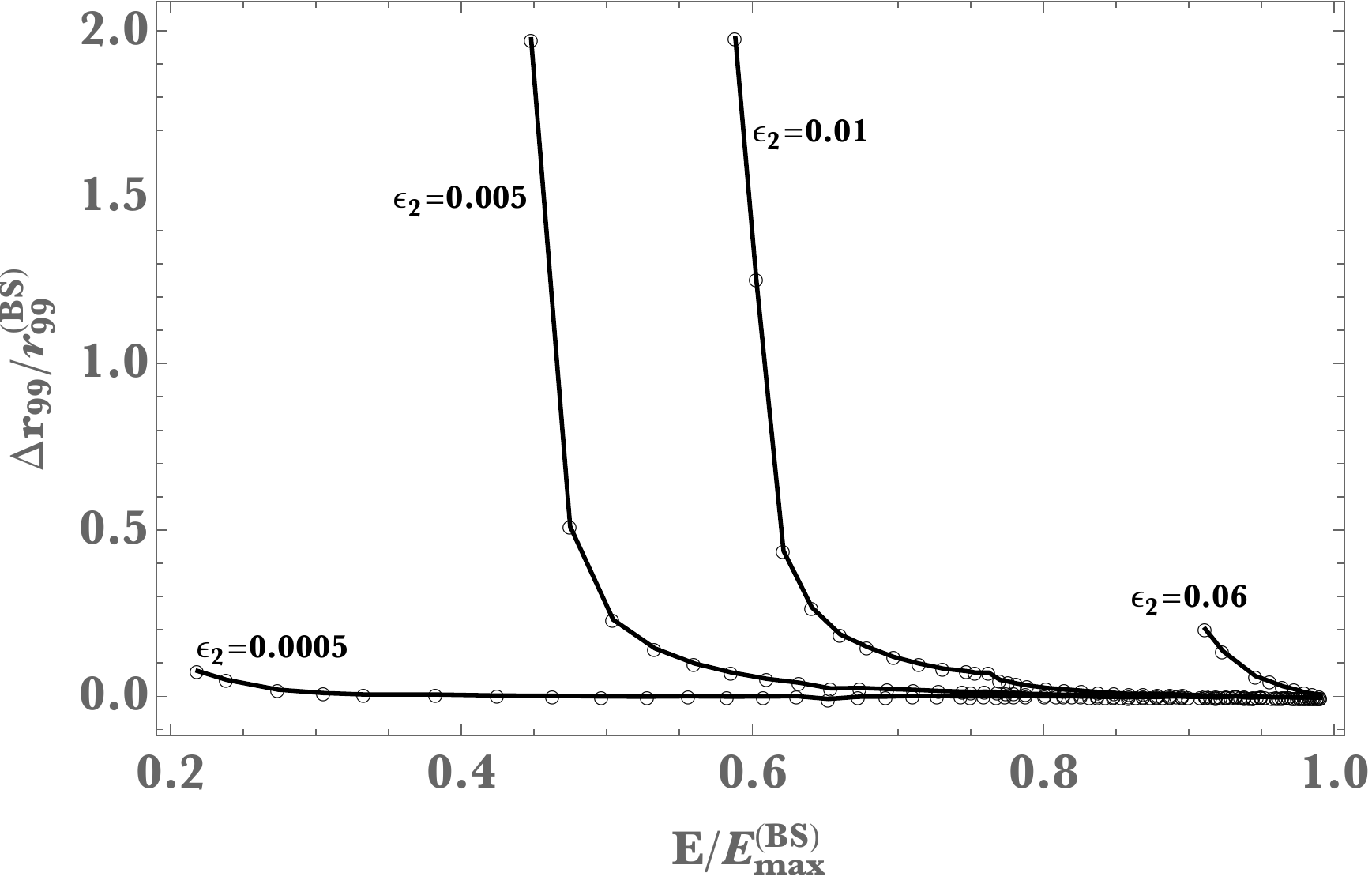}
\includegraphics[width=.4\textwidth]{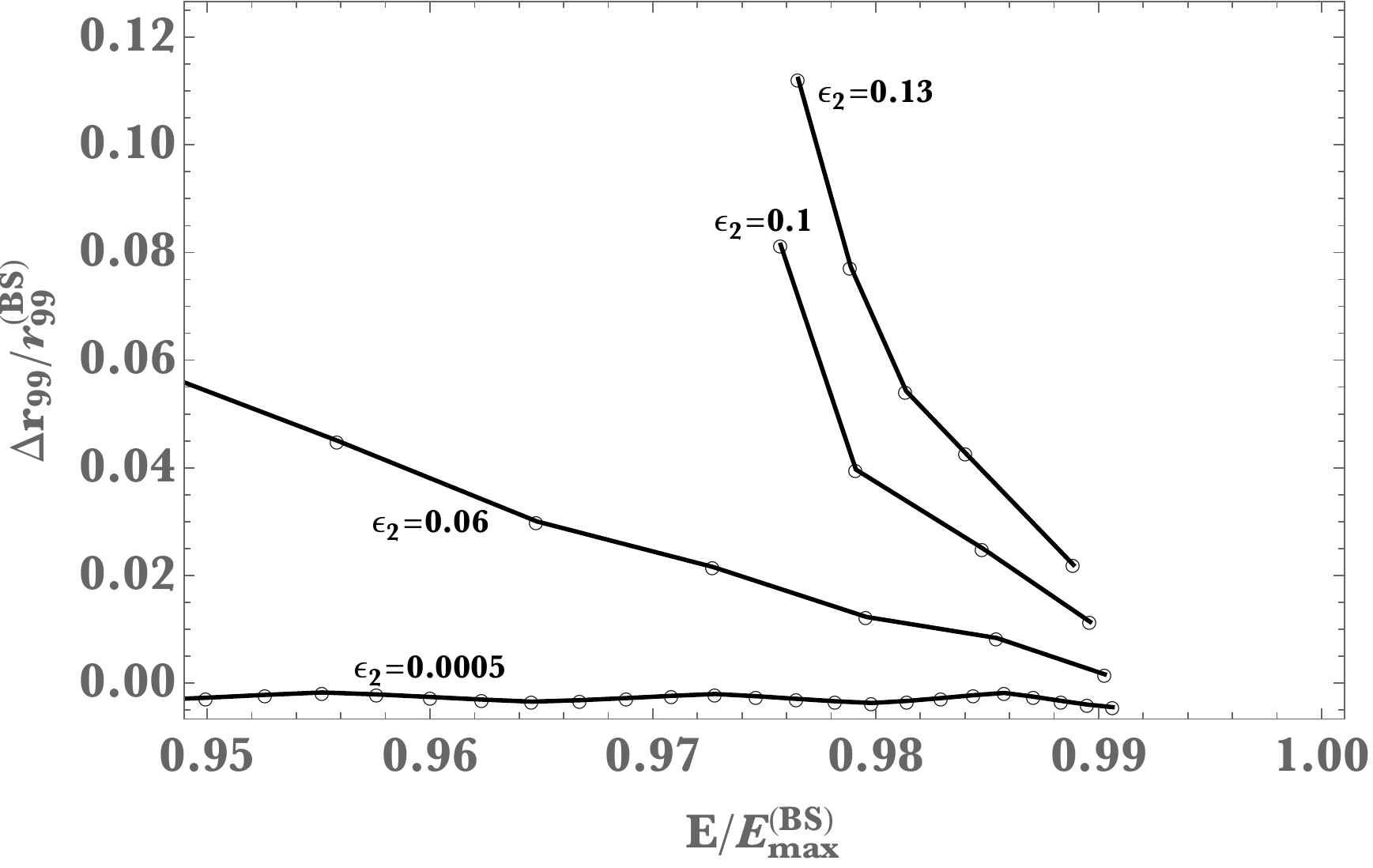}
\caption{Position at which the mass aspect function is 99\% of the total energy, as a function of the energy. Comparison is made relative to the boson star with the same energy. {\emph{Top:}} larger range of energies, in particular the line with $\epsilon_2=0.0005$ goes from frequencies $\omega_1=0.997$ to $\omega_1=0.875$. {\emph{Bottom:}} Zoom into the region of larger energies.  }
\label{fig:rM99E}
\end{figure}

To compare the size of boson stars and double-oscillators, we consider the quantity $r_{99}$, which is the spherical radius at which the mass aspect function is 99\% of the total energy \cite{Lai:2007tj}. (We take the usual spherical radius $r$ as defined just below the ansatz \eqref{ansatz} rather than the coordinate $\rho$.) In Fig.~\ref{fig:rM99E}, we again plot a relative radius $\Delta r_{\rm 99}=r_{\rm 99}/r^{(\rm BS)}_{\rm 99}-1$ as a function of energy.  We see from this figure that double-oscillators are larger and less dense objects than boson stars with the same energy. Among the solutions we have obtained, we find double-oscillators with a radius up to 200\% larger than that of the boson star with the same energy.

\textbf{Discussion --}  We have shown that boson stars are only a special case of the more general double-oscillator solutions, which we propose are part of an infinite-parameter family of multi-oscillator solutions.  The construction comes primarily from extending normal modes of a boson star to fully backreacting configurations. Our solution approach is general and can be used, in principle, to obtain any multi-oscillator solution with more frequencies.

The infinite-parameter character of the multi-oscillator family suggests that the most generic stable configuration of a complex scalar star is a multi-oscillator with some number of frequencies. Note that while we have not studied the stability of double-oscillators, the apparent stability of boson stars implies that nearby multi-oscillator solutions are also stable.

We note that there are unstable boson stars as well, whose scalar field has additional nodes in the radial direction.  Their linear perturbations include at least one unstable growing mode, along with an infinite number of stable normal modes.  Any of these normal modes can be extended to multi-oscillator solutions. When such multi-oscillators are still near the boson star and well approximated by perturbation theory, they should inherit the instability of the boson star. However, their instability remains unclear when the backreaction is much stronger. We leave the stability analysis of these solutions to future work.

Among the properties analysed, we have found that the primary and secondary frequencies of double-oscillators tend to remain close to boson stars, but their size can differ significantly even when they have the same energy.  Since double-oscillators are but a small portion of the more general multi-oscillators, we expect that the size of complex scalar stars can have considerable variation, even for a fixed energy. We expect such differences in size to be a distinguishing feature of these compact objects. In particular, the late merger dynamics of multi-oscillators might be noticeably different from those of boson stars \cite{Bezares:2017mzk,Palenzuela:2017kcg}.

We are also unconstrained by the specific matter content we have considered here. So long as there are solitonic configurations with perturbative normal modes, our methods can be used to construct multi-oscillating extensions to them.  It is natural to expect that multi-oscillating solutions built from other models, such as a real scalar field, would also exhibit similar differences in their size.

{\bf~Acknowledgements --} We thank Oscar Dias for helpful comments, and for reading an earlier version of the manuscript. R.M. acknowledges support from STFC Ernest Rutherford grant ST/M004147/1 and Univ. Southampton Global Partnerships Award 2018-19.  M.W.C. and B.W. are supported by NSERC. R.M. would like to thank the University of British Columbia for hospitality during the completion of this work.
\bibliography{refsmultiosc}{}
\end{document}